%% file: main.tex
\documentclass{article}

\PassOptionsToPackage{square, numbers}{natbib}
\usepackage[preprint]{neurips_2021}

\usepackage[utf8]{inputenc} 
\usepackage[T1]{fontenc}    
\usepackage{hyperref}       
\usepackage{url}            
\usepackage{booktabs}       
\usepackage{amsfonts}       
\usepackage{nicefrac}       
\usepackage{microtype}      
\usepackage{xcolor}         

\usepackage{caption}
\usepackage{subcaption}
\usepackage{graphicx}
\usepackage{amsmath}
\usepackage{amssymb}
\usepackage{bm}
\usepackage{subfiles}
\usepackage{multirow}
\usepackage{multicol}
\usepackage{makecell}
\usepackage{tabularx}
\usepackage{footmisc}
\usepackage{listings}
\usepackage{algorithm}
\usepackage{algorithmic}

\title{Maximizing Parallelism in Distributed Training for Huge Neural Networks}

\author{%
    Zhengda Bian\thanks{These authors have equal contributions.}\\
    National University of Singapore\\
    Singapore\\
    \texttt{zbian@comp.nus.edu.sg}\\
    \And
    Qifan Xu\footnotemark[\value{footnote}]\\
    University of California, Los Angeles\\
    Los Angeles, California, USA\\
    \texttt{QifanXu@mednet.ucla.edu}\\
    \AND
    Boxiang Wang\\
    National University of Singapore\\
    Singapore\\
    \texttt{boxiangwang059@gmail.com}\\
    \And
    Yang You\\
    National University of Singapore\\
    Singapore\\
    \texttt{youy@comp.nus.edu.sg}\\
}

\begin{document}

\maketitle

\begin{abstract}
The recent Natural Language Processing techniques have been refreshing the state-of-the-art performance at an incredible speed.
Training huge language models is therefore an imperative demand in both industry and academy.
However, the huge models impose challenges to both hardware and software.
Graphical processing units (GPUs) are iterated frequently to meet the exploding demand, and a variety of ASICs like TPUs are spawned.
However, there is still a tension between the fast growth of the extremely huge models and fact that Moore’s law is approaching the end.
To this end, many model parallelism techniques are proposed to distribute the model parameters to multiple devices, so as to alleviate the tension on both memory and computation.
Our work is the first to introduce a 3-dimensional model parallelism for expediting huge language models.
By reaching a perfect load balance, our approach presents smaller memory and communication cost than existing state-of-the-art 1-D and 2-D model parallelism.
Our experiments on 64 TACC's V100 GPUs show that our 3-D parallelism outperforms the 1-D and 2-D parallelism with 2.32X and 1.57X speedup, respectively.

\end{abstract}

\subfile{section/introduction.tex}
\subfile{section/background.tex}

\subfile{section/design.tex}


\subfile{section/evaluation.tex}

\subfile{section/conclusion.tex}

\bibliographystyle{plainnat}
\bibliography{reference}

\end{document}

%% file: section/introduction.tex
\section{Introduction}
To deal with the growing amount of data, nowadays people are squeezing the computation from GPUs or FPGAs to get larger and more powerful neural networks. 
Consequently, records are being broken in a variety of Natural Language Processing (NLP) tasks ranging from classification, question answering to translation as the language models are becoming much deeper.
In the meantime, starting from the emergence of Transformer models \cite{transformer} like BERT \cite{BERT}, the trend to increase model sizes becomes even more radical.
Since then, the Moore's Law can never reach the growth of language models.
Past several years have witnessed the creation of gigantic models, like GPT-2 \cite{gpt2} and GPT-3 \cite{gpt-3}, which are able to compose texts and synthesize images.
The amazing performance is also attracting attention from industry.
Google recently announced its Multitask Unified Model (MUM) for language understanding, LaMDA for conversation, and TPUv4 chips for model training.

The surging model size brings challenges in terms of both memory and computation.
Apart from developing chips such as GPU and TPU with higher computation capability and memory limit, activation checkpointing techniques are developed to squeeze a large model into a single device that would otherwise not be able to host such a big model. 
Moreover, ZeRO-infinity techniques \cite{zero,zero,rajbhandari2021zero} are proposed to utilize the memory of CPU or other memory device external to GPU without sacrificing the data transmission time.
Data parallelism is a dominant practice to fully utilize available HPC resources.
It distributes a large minibatch to multiple devices, where each device holds an identical model replica, and finally gathers the gradients for synchronous parameter update.
With recent optimization techniques \cite{goyal2017accurate,you2017large,you2018imagenet,you2020large}, it is now able to train very large minibatches on thousands of GPU devices.

Another branch is model parallelism, the main idea of which is to divide both computation and memory of a single neural network to multiple devices. 
Pipelined parallelism \cite{pipedream,gpipe} is proposed to split the whole model by layer, which will be executed in a pipeline fashion.
Mixed precision training is utilized to reduce both memory and computation cost.
Megatron-LM \cite{megatron} integrates both model parallelism and data parallelism by splitting the parameter tensor among model parallel groups, achieving the state-of-the-art language model.
Recently, a 2-dimensional matrix-matrix multiplication algorithm has been introduced into language model training \cite{optimus}, thus bridging the gap between traditional high-performance computing and machine learning.
In this work, we stride a step forward and propose a 3-D model parallelism technique to further harness the capability of GPU clusters.
We list our main contributions as follows.
\begin{itemize}
    \item We propose a 3-D model parallelism algorithm for linear operations, which can reach a perfect load balance, so as to provide the optimal efficiency;
    \item We use our 3-D model parallelism to implement an efficient Transformer model based on the existing PyTorch Transformer implementation and distributed communication package;
    \item We demonstrate the effectiveness of our 3-D model parallelism on 64 GPUs by comparing it with the existing 1-D and 2-D approaches. The results present the superiority of our 3-D parallelism with 2.32X and 1.57X speedup, respectively.
\end{itemize}

%% file: section/background.tex
\section{Background}
\label{sec:background}

\subsection{Transformer Language Models}
There has been a recent trend to use pretrained language models to help people train specific natural language processing (NLP) tasks, as pretrained models leverage the understanding of very large corpus (e.g. Wikipedia) to alleviate the efforts of training entire language models.
Transformer is one of the current dominant choices of pretraining language models, achieving the state-of-the-art compute efficiency and prediction accuracy.
The Transformer language model uses deep attention modules to transform a given sequential input into another sequential output.
Like LSTM, a complete Transformer model includes two parts, an Encoder and a Decoder, but each consists of attention layers rather than recurrent layers.
However, recent state-of-the-art Transformer applications, such as BERT \cite{BERT}, GPT-2 \cite{gpt2}, and GPT-3 \cite{gpt-3}, adopt more concise structures that only use necessary Encoders or Decoders on demand.

Nevertheless, training Transformer language models can be extremely computational expensive.
The recent language models advance their accuracy with a rising number of parameters (e.g. OpenAI GPT: 110M \cite{gpt1}; BERT: 340M \cite{BERT}; GPT-2: 1.5B \cite{gpt2}; GPT-3: 175B \cite{gpt-3}; Switch Transformers: 1.6T \cite{switch-tranformer}).
In the meantime, to deal with huge language corpus, there has been a growing interest in using large minibatches to reduce training time.
Therefore, efficient systems to expedite Transformers become an urgent demand for training large language models.

\subsection{Data and Model Parallelism}
Data parallelism is the most common paradigm to parallel the computation of deep neural networks.
This approach distributes the entire minibatch across multiple workers, which execute a single replica of the model and communicate with each other for synchronization at the end of each training step.
It is easy to use more devices to train larger minibatches.
By taking advantage of recent optimization techniques \cite{smith2017don,you2017large,you2018imagenet,you2020large,bytescheduler}, training speed reaches almost linear scaling with the number of devices.
However, due to the issue of generalization gap \cite{keskar2016large,hoffer2017train}, using very large minibatches requires extra efforts to guarantee the convergence performance \cite{goyal2017accurate,you2017large,you2018imagenet,you2020large}.
Besides, a practical limitation is that large Transformer models that have more than billions of parameters are basically not able to be accommodated into the memory of a single GPU device.

Model parallelism can successfully remove the above memory limitation.
There are two model parallel paradigms.
First, the layer-wise pipelined parallelism splits the entire model by layer, and execute the layers in the pipeline fashion.
Some approaches \cite{pipedream,gpipe} use their algorithms to tackle the inconsistency issue among different pipelines, so that the computation and communication time can be overlapped.
However, since each training step requires both forward and backward processing, there will be an inevitable bubble overhead, so that the compute resources cannot be fully utilized.

The second model parallel paradigm is the intra-layer parallelism, which is orthogonal to the pipelined parallelism.
It distributes the operations in each layer such as matrix multiplications and activations across multiple workers.
For example, Mesh-Tensorflow \cite{meshTensorflow}, a framework proposed by Google, provides a convenient way for users to allocate the partition of tensor dimensions. 
Then the matrix multiplications can be automatically partitioned in a single dimension and distributed to workers, without any user elaboration.
Similarly, Megatron-LM \cite{megatron} adopts a 1-D matrix partition strategy to implement the Transformer model.
It splits matrices along rows or columns, and gathers results with the all-reduce operation.
Moreover, Optimus uses the Scalable Universal Matrix Multiplication Algorithm (SUMMA) \cite{SUMMA}, a 2-D parallel strategy for matrix multiplications, which helps to reduce memory and communication costs required for executing the Transfer model.
Our approach aims to take a step further and use the 3-D parallelism to improve the performance by optimally balancing the computation, memory and communication load.

\subsection{3-D Parallel Matrix Multiplication}
In this section, we provide an outline of the 3-D parallel matrix multiplication algorithm \cite{agarwal1995three}, which is adopted in our approach.

\begin{figure}[h]
    \centering
    \includegraphics[width=0.5\linewidth]{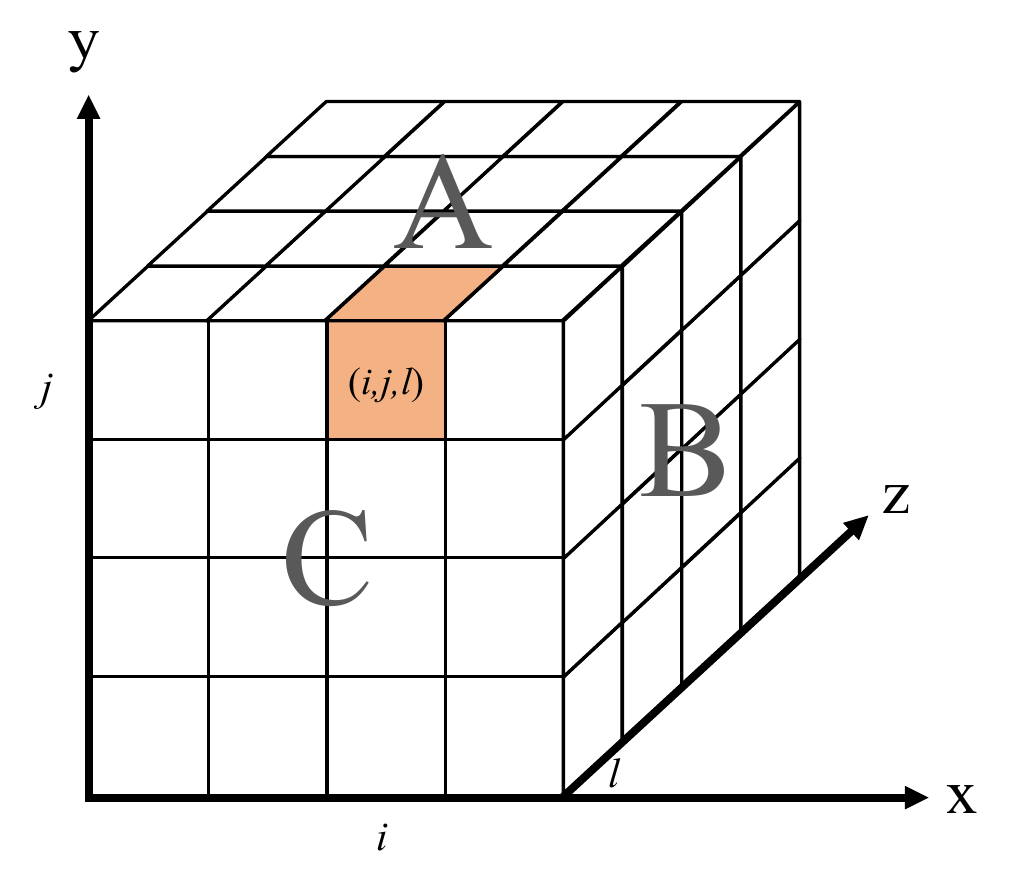}
    \caption{The structure of a 3-D processing cube with $P=p^3$ processors. The colored block $(i,j,k)$ represents an example of a single processor.}
    \label{fig:cube}
\end{figure}

This algorithm computes the multiplication between two dense matrices concurrently on $P=p^3$ processors, as illustrated in \autoref{fig:cube}, where $P$ processors can be stacked into a $p^3$ cube.
In this cube, let $A, B, C$ denote the planes, $x, y, z$ denote the directions for the planes, and $i, j, l$ denote the different indices along the directions.
We also use $A, B, C$ to represent the matrices involved in the multiplication, so that $A_{ij}, 0 \le i,j < p$ is supposed to represent a submatrix of $A$.

For simplicity, we consider an example of the 3-D matrix multiplication $C=AB$ on a $2\times2\times2$ processing cube.
We split $A$ and $B$ of size $(M,N)$ and $(N,K)$ into $2\times2$ partitions as follows, so that the sizes of each partition $A_{il}$ and $B_{lj}$ is $(M/2, N/2)$ and $(N/2, K/2)$ respectively, for $0 \le i,j < 2$.
\begin{align}
    A = \left[
        \begin{matrix}
            A_{00} & A_{01} \\
            A_{10} & A_{11}
        \end{matrix}
        \right], \quad
    B = \left[
        \begin{matrix}
            B_{00} & B_{01} \\
            B_{10} & B_{11}
        \end{matrix}
        \right].
\end{align}

\begin{figure}[H]
    \centering
    \begin{subfigure}[t]{0.25\linewidth}
        \centering
        \includegraphics[width=\linewidth]{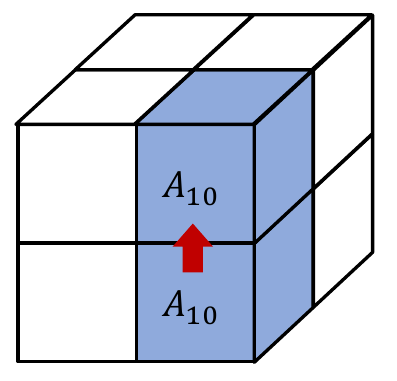}
        \caption{Broadcast $A_{il}$}
        \label{fig:3d-a}
    \end{subfigure}
    \begin{subfigure}[t]{0.25\linewidth}
        \centering
        \includegraphics[width=\linewidth]{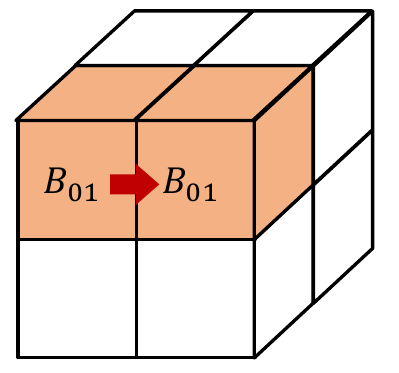}
        \caption{Broadcast $B_{lj}$}
        \label{fig:3d-b}
    \end{subfigure}
    \begin{subfigure}[t]{0.25\linewidth}
        \centering
        \includegraphics[width=\linewidth]{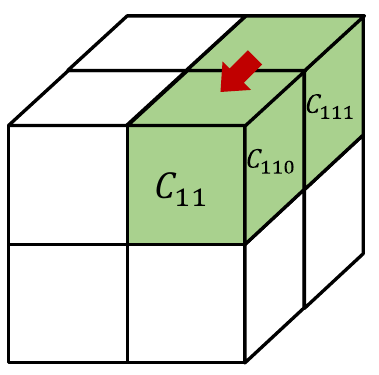}
        \caption{Reduce $C_{ij}$}
        \label{fig:3d-c}
    \end{subfigure}
    \caption{An example of the 3-D parallel matrix multiplication of size $2\times2\times2$.}
    \label{fig:3d}
\end{figure}

Each partition $A_{il}$ is stored in the processor $(i,0,l)$, while $B_{lj}$ is stored in $(0,j,l)$.
Before multiplication, we first broadcast each $A_{il}$ from $(i,0,l)$ along the $y$ direction (\autoref{fig:3d-a}), so that we have $A_il$ on each $(i,j,l)$.
Similarly, we broadcast each $B_{lj}$ along the $x$ direction (\autoref{fig:3d-b}).
Next, we can compute $C_{ijl} = A_{il}B_{lj}$ on each processor $(i,j,l)$.
Then we reduce $C_{ijl}$ to $(i,j,0)$ along the $z$ direction to derive $C_{ij}=\sum_l A_{il}B_{lj}$ (\autoref{fig:3d-c}).
Thus, we finally get $C=AB$ as
\begin{align}
    C = \left[
        \begin{matrix}
            A_{00}B_{00}+A_{01}B_{10} & A_{00}B_{01}+A_{01}B_{11} \\
            A_{10}B_{00}+A_{11}B_{10} & A_{10}B_{01}+A_{11}B_{11}
        \end{matrix}
        \right].
\end{align}

%% file: section/design.tex
\section{3-D Parallel Transformers}
\label{sec:design}

We now describe our main design and implementation of the 3-D parallel Transformer language model.

\subsection{3-D Operations}

As demonstrated in \autoref{fig:transformer}, the Transformer language model computes its hidden states from top to bottom, through multiple Transformer blocks.
In each block, the major computation in both the Self-Attention and Multi-Layer Perception (MLP) layer involves linear operations and activations.
While activation operations can be independently executed in parallel, to accelerate linear operations is the key to reduce execution time of the Transformer model.
We firstly tackle the load balancing issue of the original 3-D matrix multiplication, and then present the designs of both 3-D parallel matrix-matrix and matrix-vector operations that are used in the Transformer model.

\subsubsection{Load Balancing}
We consider the matrix multiplication $C=AB$ between the input matrix $A$ of size $(M, N)$ and the weight matrix $B$ of size $(N,K)$ on $P=p^3$ processors, which requires $MNK$ calculations in total.
An intuitive way to store the matrices is to partition each dimension of the matrices by $p$, and hold each partition $A_{il}$ on the processor $(i,0,l)$, $B_{lj}$ on $(0,j,l)$, and $C_{ij}$ on $(i,j,0)$ respectively.
In this way, we will need to broadcast each partition $A_{il}$ and $B_{lj}$ across at least $p$ processors before multiplication, and then reduce the partitions $C_{ij}$ after the multiplication, so as to guarantee the consistency among them.
Therefore, an obvious issue is that the storage is imbalanced, so that we waste a great amount of redundant memory in the processor $(i,j,l)$ if $i \neq 0$ or $j \neq 0$ or $l \neq 0$.
In the meantime, the imbalanced storage will result in imbalanced activation or element-wise operations, causing the inefficiency that the computation of such operations is not evenly distributed to all processors.

\begin{figure}[t]
\begin{minipage}{0.4\linewidth}
    \centering
    \includegraphics[width=\linewidth]{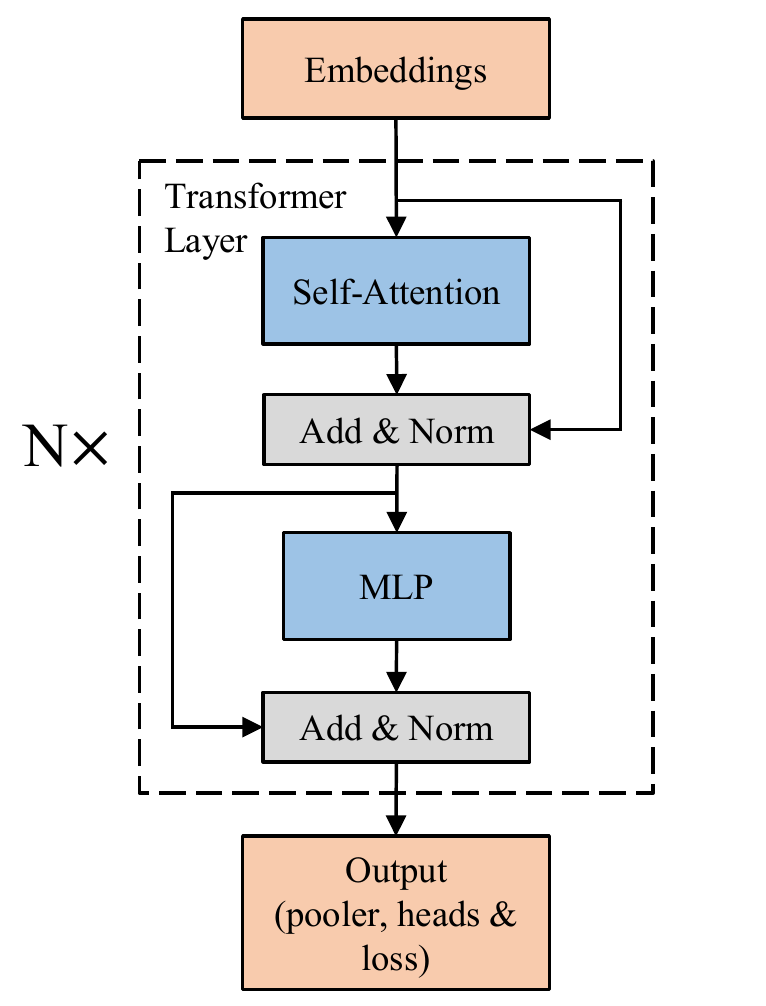}
    \caption{Architecture of the Transformer model that consists of multiple Transformer layers.}
    \label{fig:transformer}
\end{minipage}
\begin{minipage}{0.6\linewidth}
        \centering
        \begin{subfigure}[t]{0.4\linewidth}
            \centering
            \includegraphics[width=\linewidth]{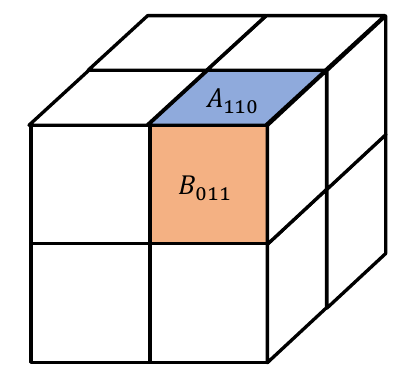}
            \caption{The input and weight submatrices on each processor.}
            \label{fig:3d2-a}
        \end{subfigure}
        \begin{subfigure}[t]{0.4\linewidth}
            \centering
            \includegraphics[width=\linewidth]{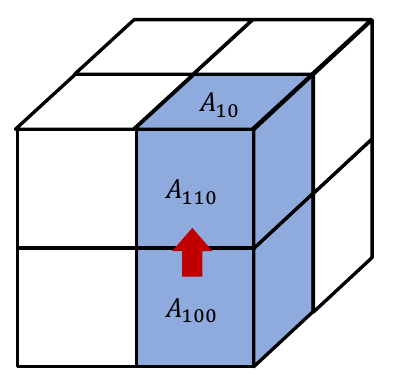}
            \caption{All-gather $A_{il}$ in the $y$ direction.}
            \label{fig:3d2-b}
        \end{subfigure}
        \begin{subfigure}[t]{0.4\linewidth}
            \centering
            \includegraphics[width=\linewidth]{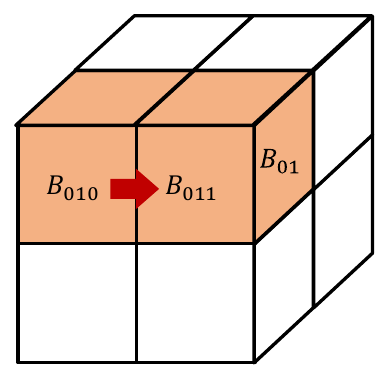}
            \caption{All-gather $B_{lj}$ in the $x$ direction.}
            \label{fig:3d2-c}
        \end{subfigure}
        \begin{subfigure}[t]{0.4\linewidth}
            \centering
            \includegraphics[width=\linewidth]{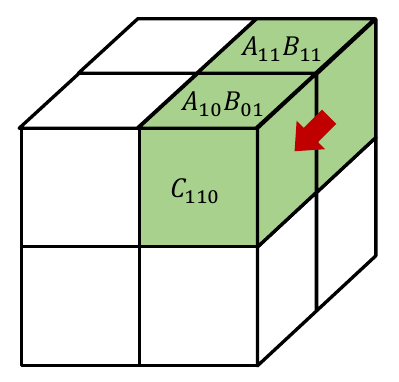}
            \caption{Compute each $C_{ijl}$ and reduce-scatter them in the $z$ direction.}
            \label{fig:3d2-d}
        \end{subfigure}
        \caption{Load-balanced 3-D matrix-matrix multiplication.}
        \label{fig:3d2}
\end{minipage}
\end{figure}

To eliminate the redundancy caused by the gap between the number of computation and memory dimensions, our insight is to distribute the matrices evenly to the cube.
More specifically, we define $m=M/p^2, n=N/p^2, k=K/p^2$.
Let each processor $(i,j,l)$ hold the partitions $A_{ijl}=A[imp+jm: imp+jm+m-1, lnp: lnp+np-1]$, $B_{lji}=B[lnp: lnp+np-1, jkp+ik: jkp+ik+k-1]$, as shown in \autoref{fig:3d2-a}.
Before the multiplication, instead of the original broadcast operation, we use an all-gather operation along the $y$ direction to copy $A_{il}=A[imp: imp+mp-1,lnp: lnp+np-1]$ across processors $(i, j, l)$ for $0 \le j < p$ (\autoref{fig:3d2-b}).
Similarly, we execute an all-gather for $B_{lj}=B[lnp: np-1, jkp: jkp+kp-1]$ (\autoref{fig:3d2-c}).
Then we compute $A_{il}B_{lj}$ on each processor, and use a reduce-scatter operation in the $z$ direction to get $C_{ij}=\sum_l A_{il}B_{lj}$ across processors $(i, j, l)$ for $0 \le l < p$.
Thus, we will finally get $C_{ilj}=C_{ij}[lm: lm+m-1, :] = C[imp+lm: imp+lm+m-1, jkp: jkp+kp-1]$ in each processor $(i,j,l)$, as shown in \autoref{fig:3d2-d}.
In overall, the memory cost per processor of our approach is supposed to be $(M/p^2)*(N/p)+(N/p)*(K/p^2)+(M/p^2)*(K/p) \approx O(1/p^3)=O(1/P)$.

\begin{figure}[h]
    \centering
    \includegraphics[width=0.5\linewidth]{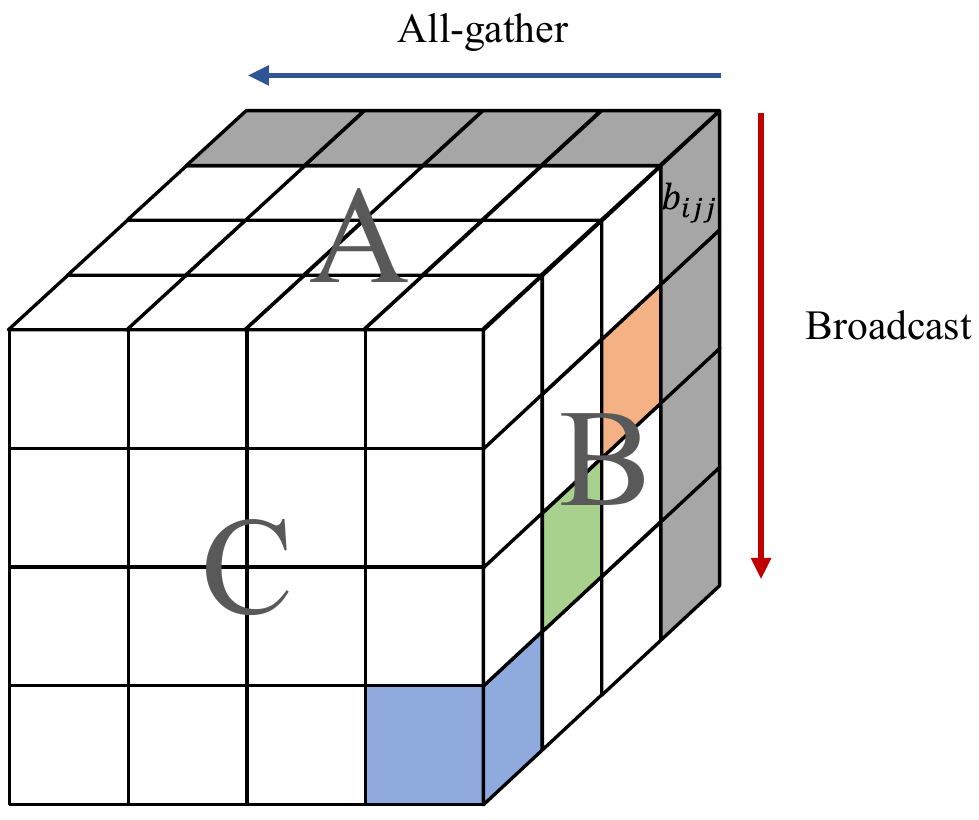}
    \caption{Balancing vector parameters for matrix-vector operations.}
    \label{fig:vector}
\end{figure}

For vector parameters that are used in matrix-vector operations, we store the vector $b$ of size $N$ diagonally on the B plane, i.e. only the processors $(i,j,l)$ holds $b_{ji} = b[jnp+in: jnp+in+n-1]$ for $j = l, 0 \le i,j,l < p$, as shown by the colored blocks in \autoref{fig:vector}.
To execute a matrix-vector operation, e.g. $A+b$, we need to broadcast $b_{li}$ in the $y$ direction, and then get $b_l$ by all-gathering them in the $x$ direction.
In contrast, to execute $C+b$, we need to broadcast $b_{ji}$ in the $z$ direction before all-gathering them.
Then we can get $A_{il}+b_l$ or $C_{ij}+b_j$ on each processor.

We need to notice that the communication directions of the input matrix $A$ and output matrix $C$ are different.
To execute the Transformer model, since the output states of each linear layer are basically the input states of the next layer, we exchange the communication directions $y$ and $z$ for the input and output states after the 3-D parallel matrix multiplication, while keep the direction $x$ for the network weights.
The matrix-vector operations are supposed to have no affect on the communication directions.

\subsubsection{Matrix-matrix Operations}
In the Transformer model, three forms of matrix-matrix multiplication, including $C=AB$, $C=AB^T$ and $C=A^TB$, are involved w.r.t the following differentiation formulas.
\begin{align}
    C=AB, \dot{A}=\dot{C}B^T, \dot{B}=A^T\dot{C}; \\
    C=AB^T, \dot{A}=\dot{C}B, \dot{B}=\dot{C}^TA; \\
    C=A^TB, \dot{A}=B\dot{C}^T, \dot{B}=A\dot{C},
\end{align}
where $\dot{X}$ denotes the gradient of the parameter $X$.
We present the algorithms for the three forms of multiplication with the optimal 3-D execution and communication efficiency as shown in the following pseudo-codes (Algorithm \autoref{alg:parallel_ab_forward}-\autoref{alg:parallel_atb_backward}).
Note that for each multiplication algorithm, other than the input matrices, we also need to specify the directions for them as well as the output matrix.

\begin{minipage}{0.45\linewidth}
    \begin{algorithm}[H]
        \caption{Forward $C=AB$}
        \label{alg:parallel_ab_forward}
        \begin{algorithmic}[1]
            \REQUIRE $A_{ijl}$, $B_{lji}$, \\
                    directions $y,x,z$ for $A,B,C$ 
            \ENSURE $C_{ilj}$
            \STATE All-gather $A_{il}$ in $y$ \\
            \STATE All-gather $B_{lj}$ in $x$ \\
            \STATE $C_{ilj} \gets A_{il}B_{lj}$ \\
            \STATE Reduce-scatter $C_{ilj}$ in $z$ \\
            \RETURN $C_{ilj}$
        \end{algorithmic}
    \end{algorithm}
\end{minipage}
\begin{minipage}{0.45\linewidth}
    \begin{algorithm}[H]
        \caption{Backward $C=AB$}
        \label{alg:parallel_ab_backward}
        \begin{algorithmic}[1]
            \REQUIRE $\dot{C}_{ilj}$, $A_{ijl}$, $B_{lji}$\\
                    directions $y,x,z$ for $A,B,C$
            \ENSURE $\dot{A}_{ijl},\dot{B}_{lji}$ \\ [0.3\baselineskip]
            \STATE $\dot{A}_{ijl} \gets \dot{C}_{ilj}B_{lji}^T$ in $z,x,y$ \\ [0.3\baselineskip]
            \STATE $\dot{B}_{lji} \gets A_{ijl}^T\dot{C}_{ilj}$ in $y,z,x$\\ [0.3\baselineskip]
            \RETURN $\dot{A}_{ijl},\dot{B}_{lji}$
        \end{algorithmic}
    \end{algorithm}
\end{minipage}

\begin{minipage}{0.45\linewidth}
    \begin{algorithm}[H]
        \caption{Forward $C=AB^T$}
        \label{alg:parallel_abt_forward}
        \begin{algorithmic}[1]
            \REQUIRE $A_{ijl}$, $B_{jli}$, \\
                    directions $y,x,z$ for $A,B,C$
            \ENSURE $C_{ilj}$
            \STATE All-gather $A_{il}$ in $y$ \\
            \STATE All-gather $B_{jl}$ in $x$ \\
            \STATE $C_{ilj} \gets A_{il}B_{jl}^T$ \\
            \STATE Reduce-scatter $C_{ilj}$ in $z$\\
            \RETURN $C_{ilj}$
        \end{algorithmic}
    \end{algorithm}
\end{minipage}
\begin{minipage}{0.45\linewidth}
    \begin{algorithm}[H]
        \caption{Backward $C=AB^T$}
        \label{alg:parallel_abt_backward}
        \begin{algorithmic}[1]
            \REQUIRE $\dot{C}_{ilj}$, $A_{ijl}$, $B_{jli}$\\
                    directions $y,x,z$ for $A,B,C$
            \ENSURE $\dot{A}_{ijl},\dot{B}_{jli}$ \\ [0.4\baselineskip]
            \STATE $\dot{A}_{ijl} \gets \dot{C}_{ilj}B_{jli}$ in $z,x,y$ \\ [0.4\baselineskip]
            \STATE $\dot{B}_{jli} \gets \dot{C}_{ilj}^TA_{ijl}$ in $z,y,x$\\ [0.4\baselineskip]
            \RETURN $\dot{A}_{ijl},\dot{B}_{jli}$
        \end{algorithmic}
    \end{algorithm}
\end{minipage}

\begin{minipage}{0.45\linewidth}
    \begin{algorithm}[H]
        \caption{Forward $C=A^TB$}
        \label{alg:parallel_atb_forward}
        \begin{algorithmic}[1]
            \REQUIRE $A_{ilj}$, $B_{ijl}$, \\
                    directions $y,x,z$ for $A,B,C$
            \ENSURE $C_{jli}$
            \STATE All-gather $A_{ij}$ in $y$ \\
            \STATE All-gather $B_{il}$ in $x$ \\
            \STATE $C_{jl} \gets A_{ij}^TB_{il}$ \\
            \STATE Reduce-scatter $C_{jli}$ in $z$\\
            \RETURN $C_{jli}$
        \end{algorithmic}
    \end{algorithm}
\end{minipage}
\begin{minipage}{0.45\linewidth}
    \begin{algorithm}[H]
        \caption{Backward $C=AB^T$}
        \label{alg:parallel_atb_backward}
        \begin{algorithmic}[1]
            \REQUIRE $\dot{C}_{jli}$, $A_{ilj}$, $B_{ijl}$\\
                    directions $y,x,z$ for $A,B,C$
            \ENSURE $\dot{A}_{ilj},\dot{B}_{ijl}$ \\ [0.35\baselineskip]
            \STATE $\dot{A}_{ilj} \gets B_{jli}\dot{C}_{ilj}^T$ in $x,z,y$ \\ [0.35\baselineskip]
            \STATE $\dot{B}_{ijl} \gets A_{ijl}\dot{C}_{ilj}$ in $y,z,x$\\ [0.35\baselineskip]
            \RETURN $\dot{A}_{ilj},\dot{B}_{ilj}$
        \end{algorithmic}
    \end{algorithm}
\end{minipage}

Our 3-D parallel matrix multiplication can evenly distribute the computational cost to all the processors, where each processor only multiplies the submatrices of size $(M/p, N/p)$ and $(N/p, K/p)$.
If we fix the problem size, the execution cost of our approach is supposed to be $(M/p)*(N/p)*(K/p) \approx O(1/p^3)=O(1/P)$.
Furthermore, compared with 1-D and 2-D approaches, each communication operation in our 3-D approach moves data across a smaller number of processors.
Each forward algorithm uses all-gather and reduce-scatter operations to move $(MN+NK+MK)/p^3$ across $p$ processors in total.
Therefore, the bandwidth cost of our approach is supposed to be $O(1/p^2)=O(P^{-2/3})$, while the latency cost is supposed to be $O(log(p))=O(log(P^{1/3}))$.

\subsubsection{Matrix-vector Operations}
The logic to execute matrix-vector operations such $C=A+b$ and $C=A*b$ is similar.
We present an example of the add operation as shown in Algorithm \autoref{alg:parallel_add_forward}-\autoref{alg:parallel_add_backward}.

For the matrix-vector multiplication, the only changes from the above algorithms are to return $C_{ijl}=A_{ijl}*b_{l}$ for the forward pass, as well as to return $\dot{A}_{ijl}=\dot{C}_{ijl}*{b}_{ijl}$ and $\dot{b}_{ijl}=\sum_{ij} \dot{C}_{ijl}*{A}_{ijl}$ for the backward pass.

\begin{minipage}{0.45\linewidth}
    \begin{algorithm}[H]
        \caption{Forward $C=A+b$}
        \label{alg:parallel_add_forward}
        \begin{algorithmic}[1]
            \REQUIRE $A_{ijl}$, $b_{ijl}$, \\
                    directions $y,x,z$ for $A,B,C$ \\ [\baselineskip]
            \ENSURE $C_{ijl}$ \\ [\baselineskip]
            \STATE Broadcast $b_{il}$ in $y$ \\ [\baselineskip]
            \STATE All-gather $b_{l}$ in $x$ \\ [\baselineskip]
            \STATE $C_{ijl} \gets A_{ijl}+b_{l}$ \\ [\baselineskip]
            \RETURN $C_{ijl}$
        \end{algorithmic}
    \end{algorithm}
\end{minipage}
\begin{minipage}{0.45\linewidth}
    \begin{algorithm}[H]
        \caption{Backward $C=A+b$}
        \label{alg:parallel_add_backward}
        \begin{algorithmic}[1]
            \REQUIRE $\dot{C}_{ijl}$,\\
                    directions $y,x,z$ for $A,B,C$
            \ENSURE $\dot{A}_{ijl},\dot{b}_{ijl}$ \\
            \STATE $\dot{A}_{ijl} \gets \dot{C}_{ijl}$ \\
            \STATE $\dot{b}_{ijl} \gets \sum_{ij} \dot{C}_{ijl}$
            \IF{$j=l$}
                \STATE Reduce-scatter $\dot{b}_{ijl}$ in $x$\\
            \ELSE
                \STATE $\dot{b}_{ijl} \gets$ null
            \ENDIF
            \RETURN $\dot{A}_{ijl},\dot{b}_{ijl}$
        \end{algorithmic}
    \end{algorithm}
\end{minipage}

By taking advantage of balanced parameter storage, we can evenly distribute the computation cost of not only the matrix-vector operations but also the activations to all the processors.
Their computation cost is supposed to be the same scale as the memory cost $O(1/P)$.

\subsection{Parallel Transformer layers}

\begin{figure}[h]
    \centering
    \begin{subfigure}[t]{\linewidth}
        \centering
        \includegraphics[width=\linewidth]{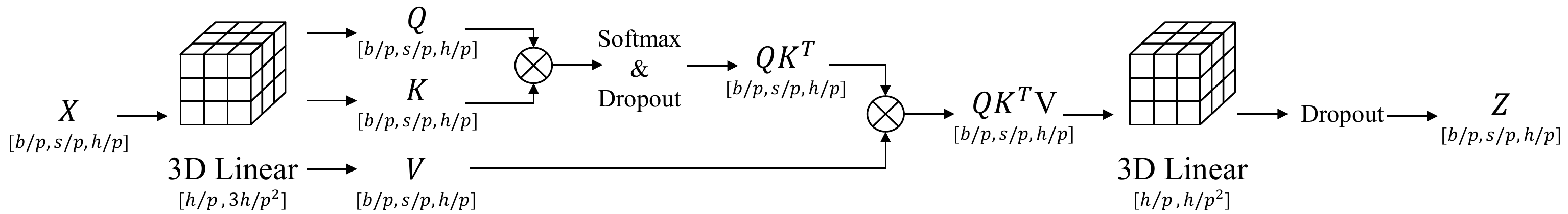}
        \caption{The Self-Attention block}
        \label{fig:attention}
    \end{subfigure}
    \begin{subfigure}[t]{0.7\linewidth}
        \centering
        \includegraphics[width=\linewidth]{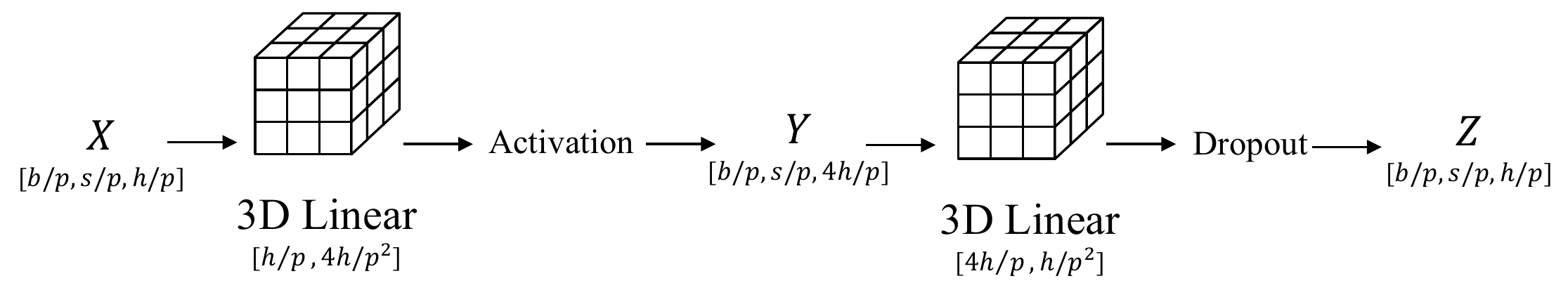}
        \caption{The Multi-Layer Perception block}
        \label{fig:mlp}
    \end{subfigure}
    \caption{Outline of 3-D Parallel Transformer layers. Each $m,n,k$ in $[m,n,k]$ represents the size of the parameter stored on each processor. The arrows present the order of computation.}
    \label{fig:layer}
\end{figure}

In this section, we describe the major implementation details of the Transformer model that uses our 3-D parallel operations.

To leverage 3-D model parallelism, we implement our new linear layer module and layer normalization module based on the 3-D parallel operations.
The 3-D linear layer flattens an input $X$ of size $[b/p,s/p,h/p]$ into size $[b*s/p^2, h/p]$ first, and then uses Algorithm \autoref{alg:parallel_ab_forward} and \autoref{alg:parallel_add_forward} to perform $Y=XW+b$ with parameters $W$ and $b$.
Thus, after a 3-D linear layer, the input and output directions should be exchanged.
The 3-D layer normalization will not affect the directions, because it only applies matrix-vector adds and multiplications with the parameters $\gamma$ and $\beta$.

We define $b,s,n,h$ as the batch size, sequence length, number of attention heads and hidden size of the Transformer model on a $p^3$ processing cube.
For ease of presentation, we do not discuss the embedding and output layers of the language model, as they usually depend on the downstream tasks.
Then our implementation can basically divided into two modules: Self-Attention blocks and MLP blocks, as illustrated in \autoref{fig:attention} and \autoref{fig:mlp}.
Both the Self-Attention and MLP block include two 3-D linear layers, so that communication directions of their inputs and outputs are supposed to be the same.
Thus, the input and output of each Transformer layer are also supposed to have the same communication direction.

Besides, in practice, it is therefore not necessary to track all the directions of 3-D parallel operations as shown in Algorithm \autoref{alg:parallel_ab_forward}-\autoref{alg:parallel_add_backward}, because we only need to exchange the input and output direction after the first linear layer of both Self-Attention and MLP blocks.
For generalizability, we index the initial input, weight and output group as 0, 1, 2.
We provide an input group index $y$ for the first linear layer, so that the output group index is supposed to be $z=1-y$.
After the layer, we take $z$ as the new input group index, and return it to $y=1-z$ after each Self-Attention or MLP block is completed.

%% file: section/evaluation.tex
\section{Experiments}
\label{sec:eval}

\subsection{Setup}
We evaluate our 3-D parallel Transformer model on TACC's Longhorn supercomputers \cite{longhorn}.
Our testbed cluster consists of up to 16 GPU servers connected by Mellanox EDR Infiniband network.
Each server has 2 20-core IBM Power 9 CPUs, 256GB memory and 4 NVIDIA V100 GPUs, so there are 64 GPUs in total.
We store the training data, model checkpoints in a Hadoop Distributed File System (HDFS) via 1Gbps Ethernet.

We compare the performance of our approach against the existing state-of-the-art 1-D \cite{megatron} and 2-D \cite{optimus} model parallel approaches.
For simplicity, we only evaluate the performance of the Transformer model itself, i.e. the consecutive Transformer layers, to highlight the efficiency of different parallelism.
We evaluate the 1-D parallelism on 8,16,36 and 64 GPUs, the 2-D parallelism on a subset of 16,36 and 64 GPUs, as well as our 3-D approach on a subset of 8 and 64 GPUs.

\subsection{Performance}

\subsubsection{Weak Scaling}

We first evaluate the weak scaling performance, where we fix the network size on each processor, and scale up the number of processors.
\autoref{tab:weak} presents the results of the weak scaling experiments, including the execution time for both forward and backward pass of the Transformer model.
We also provide the average step time of each sequence, which is computed as 
\begin{align}
    \text{Average step time} = \frac{\text{forward time} + \text{backward time}}{\text{batch size}}.
\end{align}

\begin{table}[h]
    \begin{subtable}[h]{\linewidth}
        \centering
        \begin{tabular}{c|c|c|c|c|c|c}
            \toprule
            \multirow{6}{*}{1-D \cite{megatron}} & \multirow{2}{*}{\# GPUs} & \multirow{2}{*}{\makecell[c]{Batch\\size}} & \multirow{2}{*}{\makecell[c]{Hidden\\size}} & \multirow{2}{*}{\makecell[c]{Forward\\time (s)}} & \multirow{2}{*}{\makecell[c]{Backward \\ time (s)}} & \multirow{2}{*}{\makecell[c]{Average\\ step time (s)}}\\
            & & & & & & \\
            \cline{2-7}
            & 8 & 60 & 2048 & 4.759 & 15.676 & 0.341 \\
            \cline{2-7}
            & 16 & 60 & 4096 & 12.488 & 30.894 & 0.723 \\
            \cline{2-7}
            & 36 & 40 & 6120 & 13.515 & 31.822 & 1.133 \\
            \cline{2-7}
            & 64 & 30 & 8192 & 13.915 & 32.890 & 1.560 \\
            \bottomrule
        \end{tabular}
        \label{tab:weak-1d}
    \end{subtable}
    
    \begin{subtable}[h]{\linewidth}
        \centering
        \begin{tabular}{c|c|c|c|c|c|c}
            \toprule
            \multirow{5}{*}{2-D \cite{optimus}} & \multirow{2}{*}{\# GPUs} & \multirow{2}{*}{\makecell[c]{Batch\\size}} & \multirow{2}{*}{\makecell[c]{Hidden\\size}} & \multirow{2}{*}{\makecell[c]{Forward\\time (s)}} & \multirow{2}{*}{\makecell[c]{Backward\\time (s)}} & \multirow{2}{*}{\makecell[c]{Average\\step time (s)}}\\
            & & & & & & \\
            \cline{2-7}
            & 16 & 192 & 4096 & 33.860 & 101.981 & 0.708 \\
            \cline{2-7}
            & 36 & 288 & 6120 & 54.760 & 165.850 & 0.766 \\
            \cline{2-7}
            & 64 & 384 & 8192 & 99.419 & 304.707 & 1.052 \\
            \bottomrule
        \end{tabular}
        \label{tab:weak-2d}
    \end{subtable}
    \begin{subtable}[h]{\linewidth}
        \centering
        \begin{tabular}{c|c|c|c|c|c|c}
            \toprule
            \multirow{4}{*}{~~~ 3-D ~~~} & \multirow{2}{*}{\# GPUs} & \multirow{2}{*}{\makecell[c]{Batch\\size}} & \multirow{2}{*}{\makecell[c]{Hidden\\size}}  &  \multirow{2}{*}{\makecell[c]{Forward\\time (s)}} & \multirow{2}{*}{\makecell[c]{Backward\\time (s)}} & \multirow{2}{*}{\makecell[c]{Average\\step time (s)}}\\
            & & & & & & \\
            \cline{2-7}
            & 8 & 192 & 2048 & 30.096 & 81.212 & 0.580 \\
            \cline{2-7}
            & 64 & 384 & 8192 & 79.349 & 125.037 & \textbf{0.672} \\
            \bottomrule
        \end{tabular}
        \label{tab:weak-3d}
    \end{subtable}
    \caption{Comparison of weak scaling results. The number of processors increases from 8 to 64, while the number of parameters is fixed for each approach. For each of presentation, we mainly adjust the batch size and hidden size based on the number of processors, while fix the sequence length to 512. The bolded number represents the best result.}
    \label{tab:weak}
\end{table}

We see that the 3-D parallelism has the slowest rising speed in the average step time, reaching the smallest value at the largest compute scale.
The result implies that our approach is efficient to reduce the overhead of model parallelism, as it reaches the minimum communication cost.

\subsubsection{Strong Scaling}
Next, we evaluate the strong scaling performance of different model parallelism.
This experiment aims to examine how much speedup each approach can achieve by using an increasing number of processors to execute the fix-sized problem.

The problem size and execution results are shown in \autoref{tab:strong}.
We see that our 3-D parallelism outperforms other approaches by achieving the smallest average step time on 64 GPUs, with 2.32X and 1.57X speedup over 1-D and 2-D approach, respectively.

\begin{table}[h]
    \begin{subtable}[h]{\linewidth}
        \centering
        \begin{tabular}{c|c|c|c|c|c|c}
            \toprule
            \multirow{6}{*}{1-D \cite{megatron}} & \multirow{2}{*}{\# GPUs} & \multirow{2}{*}{\makecell[c]{Batch\\size}} & \multirow{2}{*}{\makecell[c]{Hidden\\size}} & \multirow{2}{*}{\makecell[c]{Forward\\time (s)}} & \multirow{2}{*}{\makecell[c]{Backward \\ time (s)}} & \multirow{2}{*}{\makecell[c]{Average\\ step time (s)}}\\
            & & & & & & \\
            \cline{2-7}
            & 8 & 12 & 3072 & 1.470 & 5.699 & 0.597 \\
            \cline{2-7}
            & 16 & 12 & 3072 & 1.371 & 5.152 & 0.544 \\
            \cline{2-7}
            & 36 & 12 & 3072 & 1.455 & 5.414 & 0.572 \\
            \cline{2-7}
            & 64 & 12 & 3072 & 1.433 & 5.167 & 0.550 \\
            \bottomrule
        \end{tabular}
        \label{tab:strong-1d}
    \end{subtable}
    
    \begin{subtable}[h]{\linewidth}
        \centering
        \begin{tabular}{c|c|c|c|c|c|c}
            \toprule
            \multirow{5}{*}{2-D \cite{optimus}} & \multirow{2}{*}{\# GPUs} & \multirow{2}{*}{\makecell[c]{Batch\\size}} & \multirow{2}{*}{\makecell[c]{Hidden\\size}} & \multirow{2}{*}{\makecell[c]{Forward\\time (s)}} & \multirow{2}{*}{\makecell[c]{Backward\\time (s)}} & \multirow{2}{*}{\makecell[c]{Average\\step time (s)}}\\
            & & & & & & \\
            \cline{2-7}
            & 16 & 24 & 3072 & 4.680 & 13.698 & 0.766 \\
            \cline{2-7}
            & 36 & 24 & 3072 & 3.900 & 11.433 & 0.639 \\
            \cline{2-7}
            & 64 & 24 & 3072 & 3.007 & 8.920 & 0.497 \\
            \bottomrule
        \end{tabular}
        \label{tab:strong-2d}
    \end{subtable}
    \begin{subtable}[h]{\linewidth}
        \centering
        \begin{tabular}{c|c|c|c|c|c|c}
            \toprule
            \multirow{4}{*}{~~~ 3-D ~~~} & \multirow{2}{*}{\# GPUs} & \multirow{2}{*}{\makecell[c]{Batch\\size}} & \multirow{2}{*}{\makecell[c]{Hidden\\size}}  &  \multirow{2}{*}{\makecell[c]{Forward\\time (s)}} & \multirow{2}{*}{\makecell[c]{Backward\\time (s)}} & \multirow{2}{*}{\makecell[c]{Average\\step time (s)}}\\
            & & & & & & \\
            \cline{2-7}
            & 8 & 24 & 3072 & 3.249 & 9.120 & 0.515 \\
            \cline{2-7}
            & 64 & 24 & 3072 & 2.494 & 6.129 & \textbf{0.359} \\
            \bottomrule
        \end{tabular}
        \label{tab:strong-3d}
    \end{subtable}
    \caption{Comparison of strong scaling results. The problem size is fixed while the number of processors increases from 8 to 64. The bolded number represents the best result.}
    \label{tab:strong}
\end{table}

%% file: section/conclusion.tex
\section{Conclusion}
\label{sec:conclusion}
In this work, we introduce a 3-D intra-layer model parallelism algorithm for training huge neural models.
We propose a load balanced design to store and execute linear layers with minimum memory and communication cost.
By leveraging the 3-D model parallelism, we implement a 3-D parallel Transformer model and evaluate it on up to 64 GPUs.
Compared with the existing 1-D and 2-D parallelism, our Transformer model achieves a significant speedup.
However, it is still interesting to leverage our approach to train larger models on more processors.
We expect to see promising advances of our work with larger-scale compute resources in the future.